\documentclass[conference]{IEEEtran}
\usepackage{graphicx}
\usepackage{booktabs}
\usepackage{amsmath}
\usepackage{url}
\graphicspath{{figures/}}

\begin{document}

\title{A Conditional Timing Protection Level:\\
Holdover-Limited Undetected Time Error\\
Under GNSS Spoofing}

\author{\IEEEauthorblockN{Chakshu Baweja}
\IEEEauthorblockA{Ashforde O\"U \\ contact@ashforde.org}}

\maketitle

\begin{abstract}
A GNSS timing receiver under spoofing has no nominal-geometry fault to bound with
position-domain receiver autonomous integrity monitoring: the threat is a slow,
common-mode pull of served clock time that the receiver's own time-accuracy flag
need not reveal. We make three contributions of graded strength. First, a field
measurement: solving the receiver clock-solution
trajectory from the raw L1 pseudoranges, the receiver's clean dual-band position,
and broadcast ephemeris, we show that a recorded over-the-air spoof in the public
JammerTest 2024 campaign pulled a survey-grade u-blox ZED-F9P by roughly
$1.01$ ms of served time while the receiver reported a self-assessed time accuracy
of at most $51$ ns, a claim-versus-reality gap near $2\times10^{4}$. Second, an
impossibility observation: against an adversary free to choose the ramp rate, no
\emph{finite} unconditional bound on undetected time error exists under a single
self-referential clock-aided monitor, because a ramp slow enough to keep the disciplined reference
in lock-step (equivalently, below a sequential test's reference value) is never
alarmed while the integrated error grows without limit. A finite guarantee is
therefore necessarily \emph{conditional}. Third, the conditional bound itself: the
Timing Protection Level (TPL), equal to a model-free monitor's static
detectability floor plus the oscillator's coast uncertainty over the detection
latency, holds \emph{given} detection by an independent cross-satellite
consistency check that a coherent spoofer does not drive in lock-step. Each term is
a closed form over a primitive verified separately in the open Kshana simulator,
so the sum is reproducible by hand. Calibrated on the recorded attack, the holdover
budget is $114$ ns at a one-second recovery and $458$ ns even at a $60$-second
coast, between $2200$ and $8800$ times below the $1.01$ ms the receiver silently
accepted; on this slow ramp a clock-aided sequential test alone gives essentially
no protection (it alarms only after $\sim$$993\,\mu$s), whereas the model-free
consistency monitor crosses its alarm during the spoofer's ramp, minutes before the
capture. We are explicit about the boundary: the bound is calibrated on
real data but not independently validated against ground-truth field error, it
carries no aviation-style integrity-risk budget, and its long-coast value is
governed by the oscillator's long-tau red-noise floor and is reported as a swept
band rather than a single scalar. The simulator, the bound, and the calibration
example are open source under AGPL-3.0.
\end{abstract}

\section{Introduction}
Critical infrastructure increasingly takes its time from GNSS. Power grids
timestamp phasor measurements, cellular base stations align frames, financial
venues stamp trades, and data-center fabrics order events, all to a common clock
that is, in practice, a GPS-disciplined oscillator distributed over packet
networks by protocols such as IEEE 1588 PTP \cite{volpe,g8272,c37118,ptp}. The
same property that makes GNSS attractive, a free global time reference, makes it a
single point of failure now recognised at the policy level \cite{eo13905}: an
attacker who controls the signal controls the clock, and the feasibility of doing
so over the air is well established \cite{humphreys,tippenhauer}.

The defensive literature is dominated by two framings that do not, on their own,
protect a timing user. The first is \emph{spoofing detection}: a large body of
work raises an alarm when the received signal looks inauthentic, using power,
correlation-peak, angle-of-arrival, or clock-consistency cues
\cite{psiaki,humphreys,akos,dovis,jafarnia,schmidt}. Detection answers ``is
something wrong?'' but not ``how wrong can the time be before I would have
noticed?'' The second is
\emph{integrity}, formalised for aviation as a protection level: a statistical
bound on position error that, with quantified probability, is not exceeded
without an alarm \cite{brown,walter,blanch}. Protection levels are the right
\emph{shape} of guarantee, but the classical construction, receiver autonomous
integrity monitoring (RAIM) and its advanced multi-hypothesis descendants, bounds
a position error produced by anomalous ranges. A time-synchronization spoofer of
the coordinated, coherent class \cite{psiaki,tippenhauer} produces no such fault:
it can drive every satellite coherently, hold a clean four-satellite fix, and
leave the position essentially stationary while it walks the clock. A
position-domain solution with a floating clock absorbs the common-mode pull
entirely into the clock estimate, leaving no range residual, so position-domain
T-RAIM has nothing to bound. (An authentication layer such as Galileo OSNMA
\cite{osnma} attacks the problem upstream, at the signal; it is complementary to,
and does not replace, a receiver-side error bound.)

A timing user needs a bound on the time error that can be served
\emph{without detection}, given the monitor in place and the oscillator on the
bench. That quantity is the subject of this paper. We call it the Timing
Protection Level (TPL), and we make three contributions of graded strength.

\textbf{First, a field measurement of the gap (Section~\ref{sec:gap}).} Using a
recorded over-the-air spoof from JammerTest 2024 \cite{jammertest}, we solve the
receiver clock-solution trajectory from raw pseudoranges and show a real
$\sim$$1.01$ ms pull of served time while the receiver's own reported time
accuracy stayed at or below $51$ ns. The integrity flag was not merely optimistic;
it was wrong by four orders of magnitude. This is the problem the TPL exists to
address, observed in the field rather than asserted.

\textbf{Second, an impossibility observation (Section~\ref{sec:bound}).} We show
that no finite \emph{unconditional} TPL exists: against an adversary free to choose
the ramp rate, a sufficiently slow common-mode pull keeps the disciplined
reference corrupted in lock-step and is never alarmed, so the integrated undetected
error is unbounded. Any finite guarantee must be conditional, either on a
detectable ramp rate or on a cross-check that does not depend on the disciplined
reference.

\textbf{Third, the conditional bound and its real calibration
(Sections~\ref{sec:bound}--\ref{sec:honesty}).} Given detection by a model-free
cross-satellite consistency monitor, the served error is holdover-limited to the
sum of the monitor's static detectability floor and the oscillator's coast
1-sigma over the detection latency. Each term is a closed form over a primitive
verified independently: a $k$-sigma offset floor, a van Loan coast variance
calibrated to the \emph{measured} Allan deviation \cite{riley,vanloan}, and a
CUSUM time-to-alarm \cite{page}. Calibrated on the recorded attack, the bound runs
from $114$ ns at a one-second latency to $458$ ns at a $60$-second coast, between
$2200$ and $8800$ times below the $1.01$ ms. We are explicit about the limits:
the bound is calibrated, not field-validated; it carries no
integrity-risk-per-hour budget; and its long-coast value is reported as a band
swept over the oscillator's long-tau floor, not a single number.

\section{Related work}
\textbf{Spoofing feasibility and detection.} The feasibility of coordinated
over-the-air spoofing is established by Humphreys et al. \cite{humphreys} and, from
the security side, by Tippenhauer et al. \cite{tippenhauer}; Psiaki and Humphreys
\cite{psiaki}, and the surveys of Jafarnia-Jahromi et al. \cite{jafarnia} and
Schmidt et al. \cite{schmidt}, catalogue the detection cues, while Akos
\cite{akos} gives a receiver-level AGC defence and Montgomery et al.
\cite{montgomery} a multi-antenna one. These methods decide
\emph{whether} an attack is present. The TPL consumes such a detector as a
component, its detectability floor and its time-to-alarm, and turns the detector
into a quantitative error bound.

\textbf{Integrity and protection levels.} RAIM \cite{brown}, weighted RAIM
\cite{walter}, and advanced RAIM \cite{blanch} bound position error under a fault
hypothesis with a quantified missed-detection probability. The TPL borrows the
protection-level \emph{shape}, a bound not exceeded without an alarm, but retargets
it from a range fault to a coherent clock-time pull. We are explicit
(Section~\ref{sec:honesty}) that, unlike an aviation protection level, the TPL as
constructed here does not carry an integrity-risk-per-hour budget.

\textbf{Holdover and clock characterisation.} The behaviour of a free-running
oscillator over a coast interval is standard metrology: the Allan deviation
\cite{allan,riley} characterises the noise, and the phase-error variance over a
coast is obtained from the clock's power spectral densities via van Loan's
matrix-exponential integration \cite{vanloan}. The TPL's coast term is exactly this
holdover variance, evaluated over the detection latency and calibrated to the
\emph{measured} Allan deviation of the receiver under test.

\textbf{Secure time transfer.} On the theory side, Narula and Humphreys
\cite{narula} establish necessary and sufficient conditions for secure two-way
clock synchronization and show that one-way time transfer, and IEEE 1588 PTP
\cite{ptp} as specified, cannot be secured against a replaying adversary. Their
result is a formal theorem; our impossibility \emph{argument} is a weaker
receiver-side analogue in spirit: for an unconstrained-rate common-mode ramp under a
self-referential clock-aided monitor of the class above, no finite bound on
undetected time error exists.

\textbf{Timing-spoofing impact and sequential detection.} The downstream
consequence is concrete: Shepard et al. \cite{shepard} show GPS spoofing driving
phasor measurement units out of their grid timing budget. For the detector itself,
Page's CUSUM \cite{page} is the classical minimum-latency test for a change in a
monitored mean, set in the broader sequential change-detection theory of Basseville
and Nikiforov \cite{basseville}; the TPL uses its time-to-alarm to set the coast
interval. To our knowledge, no prior work combines an impossibility argument for
the unconditional case with a closed-form \emph{conditional} bound on undetected
time error, calibrated on a recorded over-the-air spoof with a self-consistency-checked
clock-solution solver.

\section{The protection-level gap, measured}
\label{sec:gap}
\textbf{Dataset and method.} JammerTest 2024 is a Norwegian live-sky campaign that
subjected commercial receivers to controlled jamming, spoofing, and meaconing at
Bleik/And\o ya, Norway ($69.275^\circ$N, $15.968^\circ$E); we use a public dataset
of survey-grade multi-band u-blox ZED-F9P recordings made at the event
\cite{jammertest}. The ZED-F9P is a high-precision RTK \emph{positioning} module
with a TCXO reference; we treat its reconstructed clock solution as the served time
a timing user would consume. The timing-grade sibling (ZED-F9T) adds receiver
clock-RAIM, an additional defence that the coherent threat model of this paper would
also have to defeat. We use scenario 2.1.1, a real over-the-air spoof, and scenario
3.1.1, a meaconing (record-and-rebroadcast) replay.

The receiver does not log its internal clock solution, so we reconstruct it. For
each epoch we solve the receiver clock-solution offset from true GPS time using the
recorded L1 pseudoranges, the receiver's own clean dual-band position held fixed,
and IGS broadcast ephemeris (IS-GPS-200 Keplerian propagation with satellite-clock,
relativistic, and Sagnac corrections) \cite{isgps200,kaplan}. For a static timing
receiver this clock-solution offset \emph{is} the served-time error a downstream
user would consume, up to the disciplining loop. We reference it to the holdover
extrapolation of the clean pre-attack segment, so that natural oscillator drift is
removed and the residual is the attack-induced pull rather than free-running bias.

We \emph{check the solver for self-consistency before trusting it}: in clean epochs
the per-satellite clock estimates agree to about $7$ to $8$ metres ($\sim$$22$ to
$26$ ns), and a deliberate time-offset sweep minimises the cross-satellite residual
at zero offset, confirming the time tag. These are internal-consistency checks, not
a comparison against an independent absolute truth; the headline pull is a
\emph{differential} excursion against the clean holdover extrapolation, so it is
robust to the constant L1-only, broadcast-ephemeris biases (uncorrected ionosphere,
troposphere, and multipath) that this solution does not model. That same uncorrected
noise inflates the clean cross-satellite spread, which makes the spread an
\emph{upper bound} on the monitor floor $\sigma_{\mathrm{mon}}$ and hence makes the
detectability-floor term conservative; the spread is a differential statistic and
does not by itself bound the oscillator's temporal instability, which the coast term
takes from the measured Allan deviation separately (Section~\ref{sec:calib}). The
$\sim$$1.01$ ms pull is four orders of magnitude above this $\sim$$22$ ns noise, so
the noise neither materially affects the gap nor undermines the floor's
conservatism. Only after this check do we read the attack segment.

\begin{figure}[t]
\centering
\includegraphics[width=\columnwidth]{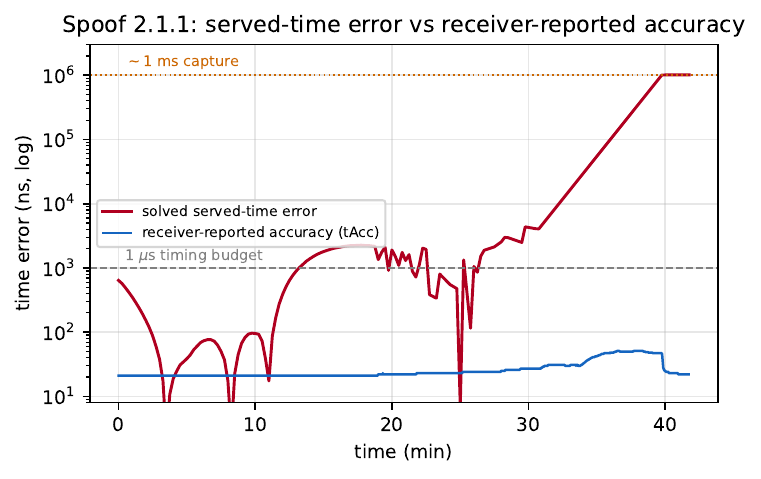}
\caption{Recorded over-the-air spoof (JammerTest 2024, scenario 2.1.1). The
served-time error solved from the raw pseudoranges (red) reaches a sustained
$\sim$1.01 ms at capture, while the receiver's self-reported time accuracy
(\texttt{tAcc}, blue) stays at or below 51 ns throughout. The dashed line is a
1 $\mu$s timing budget. The receiver's own integrity flag does not reveal the
attack.}
\label{fig:pull}
\end{figure}

\textbf{The gap.} Figure~\ref{fig:pull} shows the result. The solved served-time
error (red) rises from the clean tens-of-nanoseconds floor to a sustained
$\sim$$1.01$ ms at full spoofer capture (the solved value is not meaningful beyond
roughly three significant figures, given the $\sim$$22$ ns solver floor). Over the
same interval the receiver's self-reported time accuracy (blue, the \texttt{tAcc}
field of its navigation solution) never exceeds $51$ ns. The receiver believed it
was serving time good to $51$ ns while it was wrong by more than a millisecond, a
claim-versus-reality ratio of order $2\times10^{4}$ (Table~\ref{tab:gap}). For a
power-grid or telecom timing user with a $1\,\mu$s budget \cite{c37118,g8272},
that is the difference between in-spec and a thousandfold violation served with a
green flag.

\begin{table}[t]
\centering
\caption{Claimed versus actual integrity on the recorded attacks.}
\label{tab:gap}
\begin{tabular}{lrr}
\toprule
 & Spoof 2.1.1 & Meacon 3.1.1 \\
\midrule
Served time error (actual)        & $\sim$$1.01$ ms  & n/a \\
Receiver claimed accuracy         & $\leq 51$ ns     & $\leq 27$ ns \\
Claim-vs-reality ratio            & $\sim$$2\times10^{4}$ & n/a \\
Clean x-sat consistency floor     & $22$ ns          & $26$ ns \\
Attack x-sat consistency (peak)   & ms-scale         & $1114$ ns \\
\quad ($\sim$ floor multiple)     & $>10^{4}\times$  & $\sim$$43\times$ \\
Model-free 5$\sigma$ alarm        & yes (during ramp) & yes (at replay) \\
Clock-aided CUSUM protection      & $1\times$ ($\sim$$993\,\mu$s) & n/a \\
\bottomrule
\end{tabular}
\end{table}

\begin{figure}[t]
\centering
\includegraphics[width=\columnwidth]{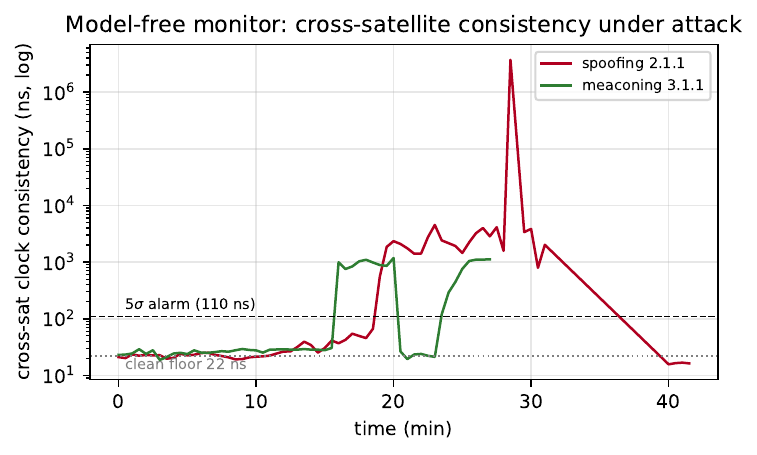}
\caption{The model-free monitor signal. Cross-satellite clock consistency
(median per bin) for the spoofing and meaconing scenarios. The clean floor is
$\sim$22 ns (dotted); under attack the binned median rises by one to two decades
and the peak reaches millisecond scale, crossing the $5\sigma$ alarm (dashed)
during the ramp. The receiver's own flag, by contrast, stays nominal.}
\label{fig:monitor}
\end{figure}

\textbf{Why the flag fails, and what catches the attack.} The attack is a
common-mode time push: a slow power ramp that drags the tracked satellites
together. Position stays essentially stationary, so there is no geometric residual,
and the receiver's \texttt{tAcc} is a formal covariance that a coherent pull leaves
small by construction, requiring no active effort from the spoofer. What a real
over-the-air spoofer \emph{cannot} keep
perfectly self-consistent is the agreement between every channel's clock solution
and an independent cross-check: a physical transmitter induces small differential
errors, so the attack is not perfectly coherent. Figure~\ref{fig:monitor} plots the
cross-satellite clock-consistency statistic for both scenarios. The clean floor is
about $22$ ns; under attack it rises by orders of magnitude to a millisecond-scale
peak, crossing a $5$-sigma ($110$ ns) alarm during the spoofer's ramp (about
$17$ min in), minutes before the $\sim$1 ms capture completes (about $21$ min in).
This model-free monitor compares satellites against each other rather than against
the (corruptible) disciplined reference, so it does not share the slow-ramp blind
spot of Section~\ref{sec:bound}: it is precisely the observable the conditional TPL
is built on, and the reason the impossibility does not leave the user defenceless.
The served error at the instant of the alarm depends on the holdover reference used
to define it (several hundred to a few thousand nanoseconds here, two to three
decades below the capture; the measured $310$ to $3700$ ns range is quantified in
Section~\ref{sec:results}); we therefore report the detection by its robust,
reference-free signature (the $5\sigma$ consistency crossing and its timing) rather
than a single reference-dependent served-error scalar.

\section{The Timing Protection Level}
\label{sec:bound}
\textbf{Threat model.} The adversary transmits a counterfeit constellation that the
victim tracks in place of the authentic signals, drives the satellites coherently,
shapes power and code phase to hold the receiver's internal integrity flag green,
and chooses the rate of the served-time ramp. The defender controls (i) a
\emph{model-free} cross-satellite consistency monitor with clean residual standard
deviation $\sigma_{\mathrm{mon}}$ and a $k$-sigma alarm, and (ii) a known
oscillator with power spectral densities $(q_{wf}, q_{rw}, q_{drift})$ for
white-FM, random-walk-FM, and drift.

\textbf{No finite unconditional bound on the served error.} A monitor that
references served time to the receiver's own coasted clock has a fundamental
limit while that clock is still being disciplined. The coast reference is only
trustworthy from the instant disciplining stops; while the loop remains slaved to
the (spoofed) signal, a ramp slower than the loop's correction is absorbed and the
self-referential residual stays small. Equivalently, a sequential test on the
\emph{rate} of that residual never alarms once the per-sample standardized
increment falls at or below the reference value $k_{\mathrm{ref}}$. An adversary
who ramps slowly enough therefore corrupts the reference in lock-step, raises no
alarm, and lets the integrated served-time error grow without bound. The escape is
not a static \emph{level} test against a \emph{frozen} reference, which catches any
offset above $k\sigma$ however slowly it was reached; but freezing the reference
requires knowing \emph{when}, which is precisely the detection the slaved monitor
cannot supply. Hence there is \emph{no finite worst-case served-time error} against
an unconstrained-rate common-mode ramp under a self-referential clock-aided monitor
alone, by which we mean any test whose decision statistic is a function of the
disciplined-reference residual standardized by a fixed clean scale; detectors with
unbounded memory or rate-of-rate statistics are a separate question we do not
address. This is why a finite guarantee must be conditional, and why the model-free
cross-check of Section~\ref{sec:gap}, which does not depend on the disciplined
reference, is the trigger the finite bound relies on: in the recorded attack it
fired on the differential leakage a coherent spoofer did not suppress.

\textbf{The conditional bound is a holdover budget.} The two monitors play
distinct roles, and the two must not be conflated. The model-free cross-satellite
monitor is the \emph{detection trigger}; how much served error accrues before it
fires is attack-dependent and reference-sensitive, and for a slow ramp is exactly
the unbounded quantity above (Section~\ref{sec:results}). The TPL
bounds what happens \emph{after} that trigger, once the receiver freezes its
reference and coasts: the residual undetected error is a static detectability floor
plus the oscillator's coast over the recovery window,
\begin{equation}
\mathrm{TPL}(\tau_{\mathrm{det}}) \;=\; \underbrace{k\,\sigma_{\mathrm{mon}}}_{\text{detectability floor}}
\;+\;
\underbrace{K\,\sigma_{\mathrm{coast}}(\tau_{\mathrm{det}})}_{\text{holdover over latency}} ,
\label{eq:tpl}
\end{equation}
so it is a \emph{holdover budget}: how far the served time can drift undetected
while the oscillator carries the clock through a detection-and-recovery latency
$\tau_{\mathrm{det}}$. The first term is the largest served-time offset that the
check against the now-frozen coast (clean residual $\sigma_{\mathrm{mon}}$, here the
$\approx$$22$ ns clock-solution consistency) does not flag at $k$ sigma. The two
terms are commensurable because both measure served-time error \emph{after} the
reference is frozen: before freezing, the reference is slaved to the spoofed signal
and a common-mode pull is invisible (the regime of the impossibility above); once
frozen at detection, the reference no longer moves with the attack, so any further
common-mode offset reappears as a residual against it. The first term is then a
genuine served-error floor, not merely a differential detection threshold, and the
second is the served-time drift the oscillator accrues over the same window; adding
them is therefore a budget on one quantity, not a mix of two. Under a Gaussian
clean-residual model,
$k=5$ would correspond to a per-test false-alarm probability of $\approx
3\times10^{-7}$ one-sided ($\approx 6\times10^{-7}$ for a two-sided consistency
alarm); we quote this only as an idealisation, because the real residual
has heavier, time-correlated tails (Section~\ref{sec:honesty}), and we do not
convert it to an integrity-risk-per-hour (that would require the full residual
distribution, the
test rate, and a target risk allocation we do not claim,
Section~\ref{sec:honesty}). The second term is the oscillator's own phase
uncertainty accumulated over the detection latency $\tau_{\mathrm{det}}$, scaled by
a coverage factor $K$; during this interval the holdover prediction, not the
spoofed signal, is the trustworthy time. The coast 1-sigma is the closed-form
integral of the clock state-space model \cite{vanloan},
\begin{equation}
\sigma_{\mathrm{coast}}(\tau) \;=\;
\sqrt{\, q_{wf}\,\tau \;+\; q_{rw}\,\frac{\tau^{3}}{3} \;+\; q_{drift}\,\frac{\tau^{5}}{20}\,},
\label{eq:coast}
\end{equation}
with the PSDs obtained from the measured Allan deviations via
$q_{wf}=\sigma_y^2(1)$ (fractional-frequency to phase), the random-walk-FM level
from the long-tau slope, and $q_{drift}$ a random-rate term (deterministic drift is
removed by the holdover reference) \cite{riley}. The TPL is a linear sum of the two
terms, reproducible by hand: at $\tau_{\mathrm{det}}=1$ s the coast is
$\sigma_y(1)\cdot 1\,\mathrm{s}=2.8$ ns on top of the $\approx$$111$ ns floor, giving
$114$ ns. We report the nominal budget at $K=1$ and state the coverage scaling
explicitly: the $K=5$ TPL at a $60$ s latency is $1844$ ns (the floor is not scaled
by $K$), still $548\times$ below the observed pull.

\textbf{Detection latency.} The latency $\tau_{\mathrm{det}}$ can be set by the
sequential detector rather than assumed. For a one-sided CUSUM with reference value
$k_{\mathrm{ref}}$ and decision interval $h$ facing a sustained standardized shift
$z$ per sample at cadence $\Delta t$, the accumulator
$S_n=\max(0,S_{n-1}+z-k_{\mathrm{ref}})$ first exceeds $h$ at
\begin{equation}
\tau_{\mathrm{det}} \;=\; \left(\left\lfloor \frac{h}{z-k_{\mathrm{ref}}} \right\rfloor + 1\right)\Delta t
\qquad (z>k_{\mathrm{ref}}),
\label{eq:cusum}
\end{equation}
and never alarms for $z\le k_{\mathrm{ref}}$, which is exactly the slow-ramp limit
above. This sets a useful $\tau_{\mathrm{det}}$ only when the clock-aided test
actually fires; on the slow ramp of the calibrated attack it does not
(Section~\ref{sec:results}), so there the latency is set by the model-free monitor's
crossing instead. In the calibration we therefore both sweep $\tau_{\mathrm{det}}$
as a design parameter (Table~\ref{tab:tpl}) and report the latency \emph{measured} by
running the monitors on the real trajectory (Section~\ref{sec:results}).

\textbf{Reproducibility of the construction.} Each term of \eqref{eq:tpl} is a
closed form over a primitive that the open Kshana simulator \cite{kshana}, an
AGPL-3.0 PNT-resilience library, verifies separately: the $k$-sigma floor against
its phase-noise model, the coast \eqref{eq:coast} against the NIST Allan stack
\cite{riley}, and the CUSUM latency \eqref{eq:cusum} against the running detector. A
reviewer can reproduce the bound by hand from the three inputs. We claim
reproducibility of the composition, which is weaker than an end-to-end validation of
its coverage, and we keep that distinction explicit.

\section{Real-data calibration}
\label{sec:calib}
\begin{figure}[t]
\centering
\includegraphics[width=\columnwidth]{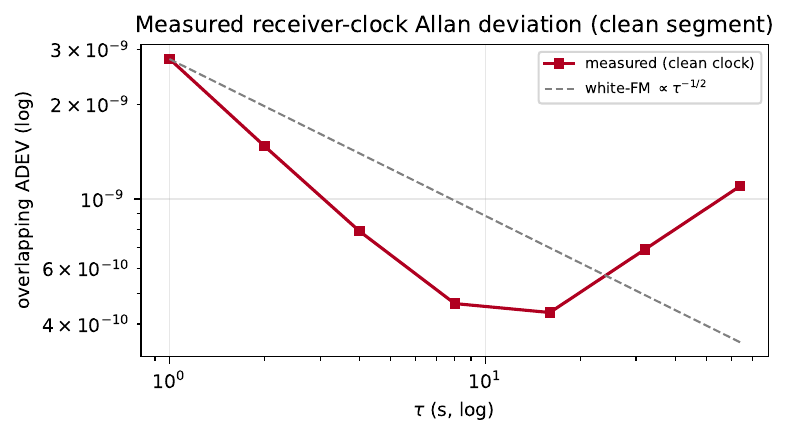}
\caption{Measured overlapping Allan deviation of the reconstructed receiver clock
on the clean segment (squares), with a $\tau^{-1/2}$ white-FM line shown only for
reference (dashed). The measured short-$\tau$ slope is steeper (near $\tau^{-0.9}$,
measurement-noise-limited), so the $1$-second value $2.8\times10^{-9}$ is used as a
conservative scalar upper bound on $\sigma_y(1)$, not as evidence of a white-FM
region; the curve turns up beyond $\tau\approx16$ s, and the long-$\tau$ red-noise
floor, below the resolution of the few-minute clean window, is swept as a band.}
\label{fig:adev}
\end{figure}

We calibrate \eqref{eq:tpl} on scenario 2.1.1. The receiver clock's overlapping
Allan deviation, measured on the clean pre-attack segment, is $2.8\times10^{-9}$ at
$\tau=1$ s (Figure~\ref{fig:adev}). Because the deviation is computed from the
pseudorange-derived clock estimate, the short-$\tau$ region is dominated by
measurement (phase) noise rather than oscillator white-FM: the measured slope from
$1$ to $4$ s is near $\tau^{-0.9}$, steeper than the $\tau^{-1/2}$ of true white-FM,
and the curve bottoms out near $\tau=16$ s before turning up. We therefore do
\emph{not} claim a clean white-FM region; we use $\sigma_y(1\,\mathrm{s})$ only as a
conservative \emph{scalar} upper bound on the oscillator's white-FM level, since
measurement noise can only inflate it. This makes the short-latency coast term
conservative, and the long-$\tau$ behaviour is carried by the swept red-noise band
(below), not by this scalar. The monitor floor is the measured clean cross-satellite
consistency $\sigma_{\mathrm{mon}}=22.1$ ns, and we set $k=5$, giving a detectability
floor of $\approx$$111$ ns (the white-FM term lifts $5\times22.1$ ns slightly). The
clean clock drift is $1618$ ns/min ($27$ ns/s), a useful holdover sanity figure.

The white-FM PSD is set from the conservative $\sigma_y(1\,\mathrm{s})$ scalar above.
The long-tau random-walk and drift PSDs cannot be observed in a clean window of only
a few minutes, so we treat them as the dominant uncertainty and sweep them
(Section~\ref{sec:honesty}). Table~\ref{tab:tpl} reports the resulting TPL and its
band versus detection latency.

\begin{table}[t]
\centering
\caption{Calibrated conditional TPL versus detection latency (scenario 2.1.1).
The band is the $\pm 1$-decade red-noise-floor sweep; the margin uses the nominal.}
\label{tab:tpl}
\begin{tabular}{rrrr}
\toprule
Latency & TPL (nom.) & Band [low, high] & Margin \\
\midrule
$1$ s   & $114$ ns & $[114, 115]$ ns  & $8849\times$ \\
$5$ s   & $119$ ns & $[118, 128]$ ns  & $8468\times$ \\
$10$ s  & $128$ ns & $[121, 158]$ ns  & $7886\times$ \\
$30$ s  & $200$ ns & $[143, 389]$ ns  & $5048\times$ \\
$60$ s  & $458$ ns & $[223, 1205]$ ns & $2208\times$ \\
\bottomrule
\end{tabular}
\end{table}

\section{Results}
\label{sec:results}
\textbf{The holdover budget is small relative to the threat.} Once the model-free
monitor has flagged the attack and the receiver coasts, the residual undetected
drift stays tightly bounded: $114$ ns at a one-second recovery and, even at a
pessimistic $60$-second coast, a nominal $458$ ns (band high $1205$ ns). At that
$60$-second row alone the budget runs from $2208\times$ (nominal) down to
$839\times$ (band high) below the $1.01$ ms a green-flagged receiver accepted; the
worst case we report anywhere is therefore $839\times$, still nearly three orders of
magnitude of margin. This
budget is computable in advance from the monitor floor and the oscillator on the
bench. It is not, and we do not present it as, a bound on how much error accrues
\emph{before} detection: that pre-detection accrual is the unbounded slow-ramp
quantity of Section~\ref{sec:bound}, and in this recording it grew into the
microsecond range before the model-free monitor's $5\sigma$ crossing (its exact
value is reference-dependent, below). The TPL bounds the recovery phase; the
model-free monitor bounds the detection phase empirically; neither alone is an
unconditional guarantee, and the paper claims neither.

\begin{figure}[t]
\centering
\includegraphics[width=\columnwidth]{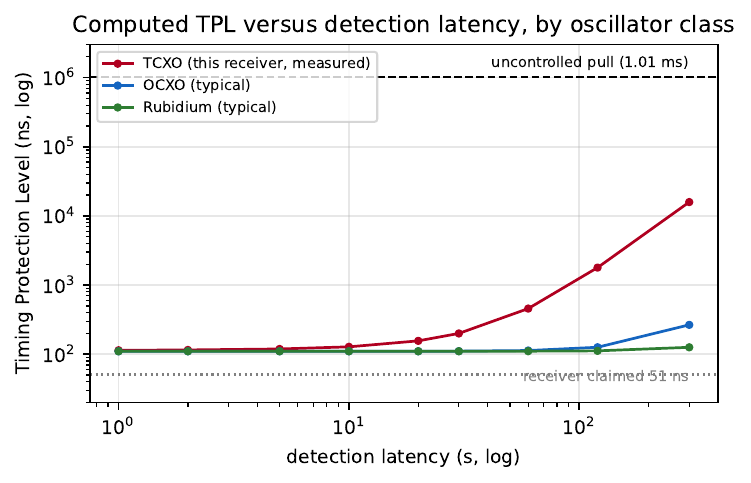}
\caption{Computed TPL versus detection latency for three oscillator classes at a
common $\approx$111 ns monitor floor. Only the TCXO is measured; the OCXO and Rb
curves are model extrapolations from class-typical PSDs. The dashed line is the
1.01 ms uncontrolled pull; the dotted line is the receiver's claimed 51 ns. Every
curve sits orders of magnitude below the uncontrolled capture across the plotted range.}
\label{fig:latency}
\end{figure}

\textbf{Oscillator class sets the achievable bound (model extrapolation).}
Figure~\ref{fig:latency} sweeps the TPL against detection latency for three
oscillator classes holding the same $\approx$$111$ ns monitor floor. Only the TCXO
curve is calibrated on measured data; the OCXO ($266$ ns at a $300$ s coast) and
rubidium ($126$ ns) curves are \emph{model extrapolations} using class-typical PSDs,
not measurements, and inherit the synthesised long-tau floor that
Section~\ref{sec:honesty} says must be measured per unit. With that caveat, the
qualitative message for a system integrator holds within the model: under the same
monitor, a better holdover oscillator buys a tighter bound at long detection
latency, and the curve says by how much.

\textbf{The two monitors, run on the real trajectory.} The clock-aided sequential
test fails exactly as the impossibility predicts. Running a CUSUM
($k_{\mathrm{ref}}=0.5$, $h=5$) on the solved clock trajectory (quadratic holdover
fit on a clean sub-segment, standardized by the robust clean sigma, with a held-out
clean window verified to raise no false alarm) does not alarm until $\sim$$560$ s
after onset, by which point $\sim$$993\,\mu$s of served error has already accrued:
a protection factor (uncontrolled-capture error divided by error-at-alarm) of
$1\times$, i.e. essentially none. The slow power ramp of
2.1.1 keeps the per-sample increment near the reference value, so the clock-aided
test never gets ahead of the attack. The model-free cross-satellite consistency
monitor of Figure~\ref{fig:monitor} does not share this blind spot: it crosses its
$5\sigma$ alarm during the ramp (about $17$ min in, against a capture completing
around $21$ min), while the served error is still two to three decades below the
$\sim$1 ms capture. The served error at that crossing is sensitive to the holdover
reference used to define it (we measured $\sim$$310$ to $\sim$$3700$ ns across
defensible quadratic and linear fits), which is why we headline the robust
reference-free signature, the $5\sigma$ crossing and its timing, and not a single
served-error scalar; we report the sensitivity rather than tuning for the best
number.

\textbf{The slow-ramp limit, quantified.} The slow-ramp limit of
Section~\ref{sec:bound} has a concrete calibrated threshold. A clock-aided CUSUM
detects a sustained common-mode ramp only when its per-sample standardized
increment exceeds the reference value, so the minimum detectable ramp rate is
$\dot r_{\min} = k_{\mathrm{ref}}\,\sigma/\Delta t$, where $\sigma$ is the scale that
standardizes the test. Taking $\sigma$ equal to the clean clock-solution consistency
$\sigma_{\mathrm{mon}}=22.1$ ns (the two clean scales agree to within their
estimation error here) and $k_{\mathrm{ref}}=0.5$ at $1$ Hz gives
$\approx 11$ ns/s: a coherent pull slower than that is invisible to the
clock-aided test indefinitely, yet still accrues $\sim$$955\,\mu$s per day, far
outside any infrastructure timing budget over a sustained operation. This is the
impossibility made numeric, and the recorded attack confirms it: the clock-aided
CUSUM gave $1\times$ protection (Section above). The protection therefore cannot
rest on the clock-aided test alone. The model-free consistency monitor does not
share the limit, because it keys on the differential inconsistency that an
imperfectly coherent spoofer leaks rather than on the corruptible reference; in the
recorded attack it crossed its $5\sigma$ alarm during the ramp and reached a
millisecond-scale peak (and a $1114$ ns peak on the meaconing replay). Its own
missed-detection probability against a coherence-optimising adversary is, however,
uncharacterised, and that, not the clock-aided floor, is the assumption the
guarantee rests on (Section~\ref{sec:honesty}).

\section{Limitations and honesty boundary}
\label{sec:honesty}
We state the limits of the claims precisely.

\textbf{No integrity-risk budget.} An aviation protection level is tied to a target
integrity risk (for example $10^{-7}$/hr) from which the $\sigma$-multipliers are
derived. The TPL as constructed here is not: $k=5$ sets a per-test false-alarm
probability under a Gaussian assumption, but we do not assert a residual
distribution in the tails, a test rate, or a missed-detection probability for the
\emph{detection event} on which the bound is conditioned. The TPL is a calibrated,
reproducible engineering bound, not a certified integrity guarantee, and we do not
use the word ``certified'' for it.

\textbf{The model-free monitor's missed-detection is the critical assumption.}
The entire finite guarantee is conditional on the cross-satellite monitor detecting
the attack, and that monitor's missed-detection probability against a
coherence-optimising adversary is not characterised here. In the recorded attack the
spoofer was imperfectly coherent and leaked a large differential inconsistency, so
the monitor fired; we say it \emph{does not} suppress that leakage in this recording
and in current spoofer hardware, not that it \emph{cannot}. A multi-antenna or
carefully calibrated transmitter, the inverse of the antenna-diversity defence of
Montgomery et al. \cite{montgomery}, could drive the differential statistic down at
any ramp rate, which would move the unbounded slow-ramp regime up to the model-free
monitor. Characterising that monitor's missed-detection floor against an optimising
adversary, and its slowest reliably detectable ramp, is the necessary next step.

\textbf{The clock-aided test alone gives no finite bound.} As Section~\ref{sec:bound}
shows, a sufficiently slow ramp evades the clock-aided test entirely;
Section~\ref{sec:results} quantifies its floor at $\approx 11$ ns/s and confirms it
on the recording (a $1\times$ protection factor). The clock-aided sequential test
is therefore a latency-tightener where it applies, not a stand-alone defence.

\begin{figure}[t]
\centering
\includegraphics[width=\columnwidth]{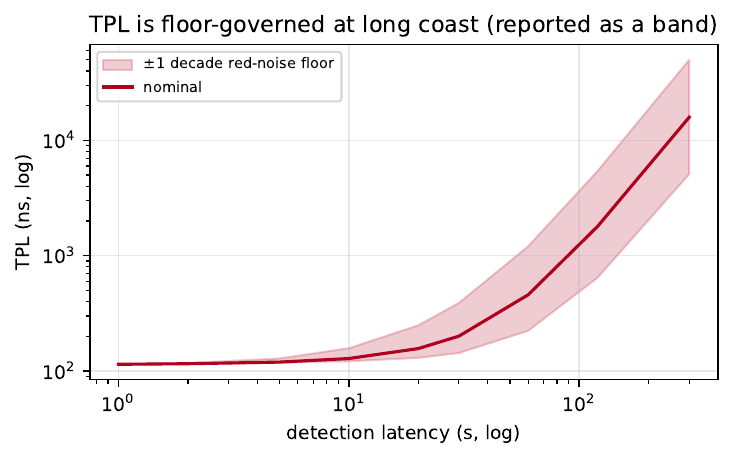}
\caption{The TPL is floor-governed at long coast. Sweeping the long-tau red-noise
floor $\pm$1 decade in PSD (shaded) leaves the short-latency bound essentially
fixed but widens it materially at long detection latency. The bound must be
reported as a band; the critical input is the oscillator's measured long-tau
stability.}
\label{fig:band}
\end{figure}

\textbf{The bound is floor-governed at long coast.} The long-tau red-noise PSDs
$(q_{rw}, q_{drift})$ are not observable in a few-minute clean window, so the coast
term at large $\tau_{\mathrm{det}}$ inherits their uncertainty. We report the TPL as
a band swept $\pm1$ decade in PSD over the red-noise floor (Table~\ref{tab:tpl},
Figure~\ref{fig:band}); $\pm1$ decade is a working choice, and a unit's true
random-walk-FM and drift can span more than a decade across temperature, so the
band is a guide, not a guarantee. A defensible per-unit figure requires the
oscillator's \emph{measured} long-tau stability; the band, not the nominal scalar,
is the appropriate object to report.

\textbf{Verified primitives are not a validated bound.} The TPL is a modelled
composition of \emph{verified} primitives, where ``verified'' means each closed form
was checked against its own reference model, not against ground truth. Its inputs are
measured from a real recorded attack, and each component term is checkable in
isolation, but the end-to-end bound has not been \emph{validated} against an
independent ground-truth time error across many attacks and receivers.
We claim a calibrated, reproducible construction, not field-proven coverage.
Demonstrating empirical coverage across a receiver and attack population is the
necessary next step and is outside this paper.

\textbf{Reconstructed clock solution, not steered PPS.} We solve the receiver's
clock-solution offset, not its steered hardware PPS output, which the dataset does
not expose. For a static timing receiver the two coincide up to the disciplining
loop; a direct PPS measurement would settle the small residual difference.

\textbf{Single campaign, two scenarios.} The calibration rests on one campaign, one
receiver, and two attack scenarios. The $\sim$$1.01$ ms gap and the monitor
behaviour are real and reproduced by two independent estimates, but generalisation
to other receivers, oscillators, and attack waveforms is asserted by construction,
not yet measured.

\section{Conclusion}
A GNSS timing receiver's self-reported integrity is not a safety bound: on a
recorded over-the-air spoof, a survey-grade receiver served a $\sim$$1.01$ ms time
error while claiming $51$ ns, a gap of order $2\times10^{4}$. We argued that no
finite unconditional bound on undetected time error exists against an
unconstrained-rate common-mode ramp under a self-referential clock-aided monitor, so
the achievable guarantee is necessarily conditional, and we gave that conditional
bound: the Timing Protection Level, a
holdover budget equal to a detectability floor plus the oscillator's coast over the
detection-and-recovery latency, each a closed form over a separately verified
primitive. Calibrated on the recorded attack, the budget is $114$ ns at a
one-second recovery and $458$ ns (band $[223,1205]$ ns) at a $60$-second coast,
thousands of times below the uncontrolled capture. The recording bears out the structure:
the clock-aided sequential test alone gave essentially no protection on the slow
ramp ($\sim$$993\,\mu$s, $1\times$), while the model-free consistency monitor
crossed its alarm during the ramp, before the uncontrolled capture. We are explicit that the
bound is calibrated rather than field-validated, carries no integrity-risk budget,
is conditioned on a model-free monitor whose missed-detection against an optimising
adversary we do not characterise, and must be read as a band at long coast. The
Kshana simulator \cite{kshana}, the bound, the calibration example, and the figure
pipeline are open source under AGPL-3.0 so that the construction can be reproduced
and the next steps, the monitor's slowest detectable ramp and empirical coverage
across receivers and attacks, can be carried out in the open.

\section*{Reproducibility}
The TPL primitives, the CUSUM detector, the Allan-deviation stack, and the
calibrated example (\texttt{cargo run --example tpl\_jammertest}) are part of the
open Kshana simulator \cite{kshana} under AGPL-3.0 and reproduce Table~\ref{tab:tpl}
exactly. The clock-solution solver and the figure pipeline that derive the real-data
figures from the recordings accompany the paper rather than the core library. The
JammerTest 2024 recordings are distributed by their authors under GPL-3.0-or-later
\cite{jammertest} and are not included here; only the scalars solved from them, with
the verification described in Section~\ref{sec:gap}, are reproduced.

\end{document}